\documentclass[sigconf]{acmart}

\AtBeginDocument{%
  }

\newcommand{\modelname}{\textsc{ECHO}}

\newcommand{\FigOneScale}{1.0} 
\newcommand{\FigThreeScale}{0.95} 
\newcommand{\FigFourScale}{1.0} 
\newcommand{\FigFiveScale}{1.0} 
\newcommand{\FigSixScale}{0.85} 
\newcommand{\FigSevenScale}{1.0} 
\newcommand{\FigEightScale}{0.75} 
\newcommand{\FigNineScale}{1.02} 

\usepackage{algorithm}
\usepackage{algpseudocode}

\setcopyright{acmlicensed}
\copyrightyear{2018}
\acmYear{2018}
\acmDOI{XXXXXXX.XXXXXXX}
\acmConference[Conference acronym 'XX]{}{June 03--05,
  2018}{Woodstock, NY}
\acmISBN{978-1-4503-XXXX-X/2018/06}

\begin{document}

\title{ECHO: Efficient Cascade with Homogeneous Operators for Distillation-Guided 100K-Sequence Modeling}


\author{
Wenhao Li,
Qinglin Jia,
Chenghou Jin,
Zihong Xian,
Jiezhou Ji, 
Yanzhi Liu, 
Zezheng Wang,\\
Bo Chen,
Ruiming Tang,
Chaoyi Ma$^{\dagger}$,
Zhaojie Liu$^{\dagger}$, 
Wenwu Ou$^{\dagger}$
}

\affiliation{%
  \institution{Kuaishou Technology}
  \city{Beijing}
  \country{China}
}

\email{{liwenhao05, machaoyi03, zhaotianxing, luocheng10}@kuaishou.com}


\renewcommand{\shortauthors}{Li et al.}

\begin{abstract}

Modeling lifetime user behavior sequences is essential for capturing long-range preferences in industrial recommendation, yet 100K-length user sequences make target-aware ranking prohibitively expensive. Search-based methods that retrieve from raw sequences incur candidate-dependent costs that grow with sequence length, whereas latent compression methods reduce computation at the expense of item-level addressability. We propose \modelname{} (\textbf{E}fficient \textbf{C}ascade with \textbf{H}omogeneous \textbf{O}perators), a framework that balances efficiency and fine-grained retrieval by retaining a subset of original event records per user via a Distilled Selector (DS) and retrieving a smaller set of target-relevant records per candidate via a Target-aware Fine-grained Retriever (TFR). DS is supervised by a candidate-aware full-history teacher via per-target distillation, enabling target-aware user-level selection without incurring full-history computation for every candidate. The retriever and target-attention scorer share the teacher's learned relevance parameters, ensuring consistent relevance estimation across the cascade. At a sequence length of \(L=100\mathrm{K}\), \modelname{} reduces online sequence-branch FLOPs by approximately \(160\times\) compared with a representative end-to-end model. Three deployment optimizations---common user inputs, compact record encoding, and key-side projection absorption---further reduce memory, storage, and data-movement overhead. Experiments on KuaiRand-1K and an industrial-scale short-video platform demonstrate relative GAUC improvements of \(0.113\%\)–\(1.520\%\) over state-of-the-art (SOTA) baselines across multiple tasks. Furthermore, online A/B testing in two production scenarios, Kuaishou and Kuaishou Lite, shows that \modelname{} improves app usage time by \(+0.240\%\) and \(+0.373\%\), respectively.

\end{abstract}

\received{20 February 2007}
\received[revised]{12 March 2009}
\received[accepted]{5 June 2009}

\keywords{Recommender Systems, User Behavior Modeling, Knowledge Distillation}

\maketitle

\section{Introduction}

In industrial-scale short-video recommendation, user interactions continuously accumulate into long behavior sequences that provide rich and diverse evidence for modeling users' evolving preferences~\cite{deng2025onerec,he2023survey,pan2026survey}.
For active users on Kuaishou, these sequences can easily reach $10^4$ events within a single month. Such histories capture not only recent intent but also stable preferences, recurring interests, and infrequent interests that may be difficult to identify from short behavioral windows~\cite{zhang2025onetrans,li2026iat,xu2025climber}. 
Effectively exploiting these long-term behavioral signals has therefore become an important direction for improving ranking performance in large-scale industrial recommendation~\cite{chen2026uniformer,li2026teacher,wang2026sample,li2026action,li2024scene}.

However, directly modeling raw behavior sequences during ranking is prohibitively expensive~\cite{ding2026bending,chen2025massive}. A ranking request typically needs to score multiple candidate items for the same user, and target-aware history modeling must evaluate historical behaviors with respect to each individual candidate~\cite{hyformer,mixformer}. As user behavior sequences grow to $10^5$ events or more, this candidate-dependent computation becomes prohibitively expensive under strict industrial latency and throughput constraints.

To make long-history modeling practical, a line of search-based methods adopts a two-stage paradigm. A lightweight General Search Unit (GSU) first retrieves a small subset of target-relevant behaviors from the long history, and an Exact Search Unit (ESU) then applies more expressive target attention over the retrieved behaviors. Representative methods such as SIM and TWIN~\cite{pi2020search,chang2023twin} are attractive because they preserve the original item records selected from the user's history. The downstream ranker can therefore still perform fine-grained target-to-item matching and trace its evidence back to concrete historical behaviors. However, the GSU must still search over the original behavior sequence for each candidate, causing its online computation cost to grow with the sequence length. Consequently, scaling search-based methods to histories containing around 100K events remains challenging under real-world industrial ranking constraints.

Another line of work reduces the cost of ultra-long history modeling through compression. These methods summarize user histories into a compact set of latent states, sketches, or interest vectors, and perform target-conditioned reasoning over the compressed representations~\cite{wang2025transformers,zhou2026sita}. SequenceO1~\cite{guan2026sequenceo1}, for example, scales to sequences of up to 100K events using a fixed-size user-only sketch and cache reuse, alongside a recent-history branch. Its sketch branch bounds candidate-dependent computation without repeatedly scanning the full history. However, compressing diverse behaviors into a finite-capacity representation before observing the target may blend interest-specific details and attenuate sparse behavioral evidence. This is a particular concern for users with dispersed interests or long-tail targets whose relevance depends on only a few historical interactions. Within the compressed sketch, such interactions are no longer individually searchable. Retaining original records offers an alternative, allowing downstream matching to directly access the concrete behaviors that survive selection.

A natural way to combine the advantages of the two directions is to first compress the search space and then retrieve from it. TWIN V2~\cite{si2024twin} demonstrates this paradigm by compressing lifelong behaviors in a target-independent manner, retaining a single representative item for each cluster, and subsequently performing online retrieval with cluster-aware target attention. This design extends the searchable range beyond what a direct GSU can efficiently handle. However, because the first-stage compression is target-independent and retains only one item per cluster, the original behaviors grouped into the same cluster are no longer independently accessible to the online retriever. This motivates us to explore a complementary form of compression that reduces the search space while preserving the original event records.

Our goal is to shorten the history before target-aware ranking while preserving as much prediction-relevant evidence of user interests as possible and keeping retained behaviors individually searchable. We pursue this goal by selecting original event records with their item identities and side features intact, guided by relevance signals learned from downstream prediction rather than solely by recency heuristics or offline structure.

We propose \modelname{}, a distillation-guided framework for lifetime behavior modeling with item-level evidence. Its name stands for \textbf{E}fficient \textbf{C}ascade with \textbf{H}omogeneous \textbf{O}perators. During training, a user-level Distilled Selector (DS) retains $K$ original event records from each user's history and materializes them for reuse across requests and candidates. For each target item, a Target-aware Fine-grained Retriever (TFR) searches only the retained records and selects at most $M$ behaviors for final interest aggregation. With $K=10{,}000$ and $M=100$ in our production setting, candidate-dependent online computation is confined to the retained history rather than the full 100K-event sequence. The selector is supervised by a candidate-aware full-history teacher trained with the prediction objective. Following TWIN's scoring-consistency strategy~\cite{chang2023twin}, the retriever and target-attention scorer also share the teacher's learned relevance parameters, aligning selection supervision with downstream matching.

The main contributions of this work are summarized as follows:
\begin{itemize}
    \item We propose \modelname{}, a distillation-guided cascade that combines user-level selection with target-aware retrieval to efficiently model lifetime user behaviors. By distilling target-aware relevance from a full-history teacher and retaining original event records with their item identities and side features intact, \modelname{} enables candidate-specific matching while preserving item-level addressability.
    \item We reduce online sequence-branch FLOPs by approximately $160\times$ relative to a representative end-to-end model at $L=100\mathrm{K}$, and describe three deployment optimizations that reduce memory, storage, and data-movement costs.
    \item Extensive experiments on KuaiRand-1K and an industrial-scale short-video platform demonstrate relative GAUC improvements of $0.113\%$--$1.520\%$ over state-of-the-art baselines across multiple tasks. Online A/B testing in Kuaishou and Kuaishou Lite further shows improvements of $+0.240\%$ and $+0.373\%$ in app usage time, respectively.
\end{itemize}

\section{Related Work}

\begin{figure*}[t]
    \centering
    \scalebox{\FigOneScale}{\includegraphics[width=0.8\textwidth]{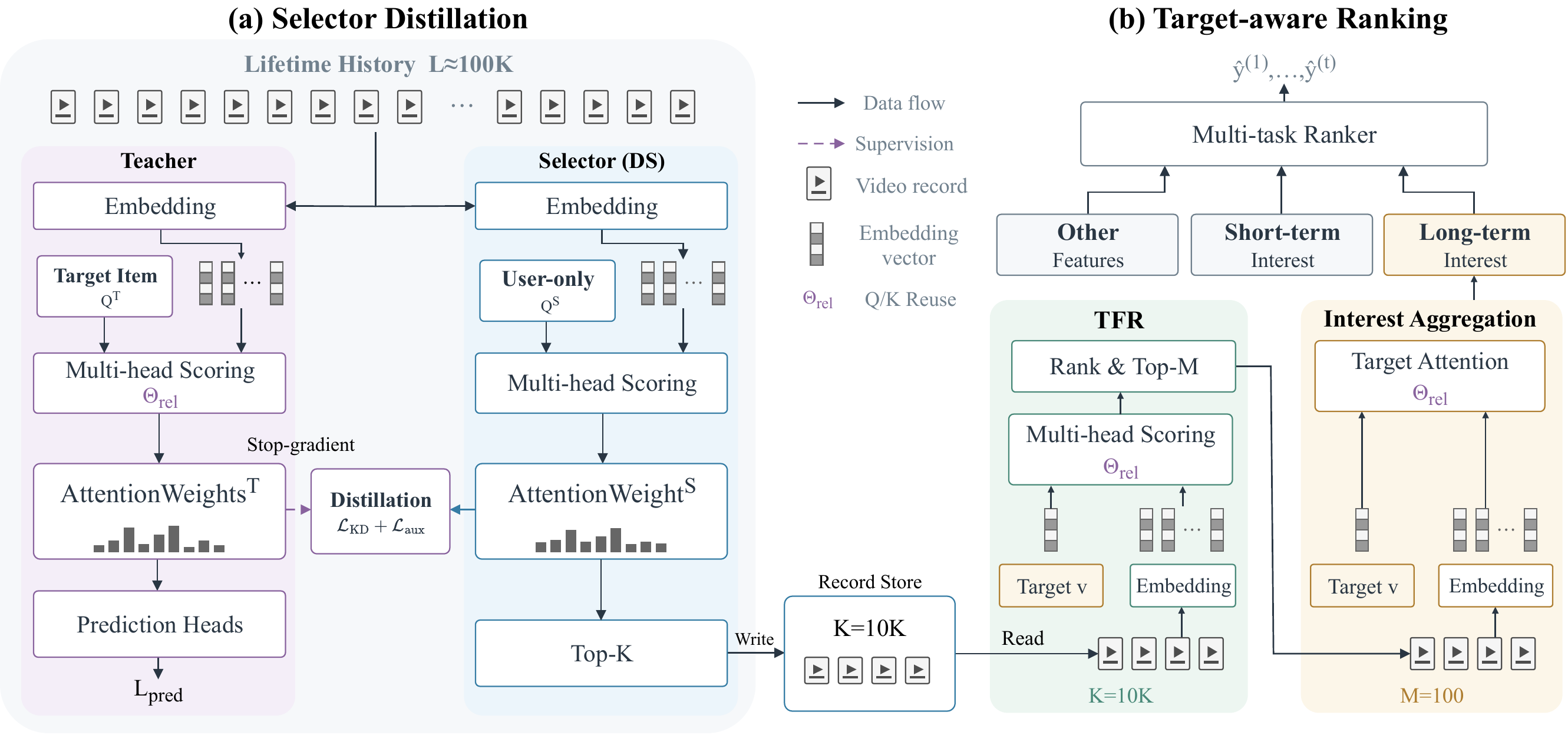}}
    \caption{Overview of \modelname{}. (a) A full-history teacher distills a user-only selector that stores $K$ original records. (b) The ranking graph retrieves $M$ target-relevant records for interest aggregation and multi-task prediction. 
    }
    \Description{Panel (a) shows a full-history teacher supervising a user-only selector, which writes Top-K original records to a shared store. Panel (b) shows target-aware retrieval, interest aggregation, and multi-task ranking. Matching Theta-rel labels denote query/key parameter reuse. The teacher and selector do not run online.}
    \label{fig:architecture}
\end{figure*}

Existing long-sequence models can be broadly categorized as end-to-end, retrieval-based, compression-based, and hybrid methods~\cite{zhou2026survey}.


\textbf{End-to-end long-sequence modeling.} These methods directly extend the ranker's capacity for longer histories. LONGER~\cite{chai2025longer} reduces sequence length through cross-attention before deeper self-attention, STCA~\cite{guan2026make} combines stacked target-to-history cross-attention with batching and length extrapolation, and HSTU~\cite{zhai2024actions} improves efficiency through specialized attention operators and sample organization. However, because these methods directly process the input history, their computational cost grows approximately linearly with sequence length, making them difficult to scale to 100K-level histories.


\textbf{Retrieval-based long-sequence modeling.} These methods retrieve target-relevant records before fine-grained interest modeling. SIM~\cite{pi2020search} establishes this paradigm, ETA~\cite{chen2021end} accelerates retrieval through approximate matching, and TWIN~\cite{chang2023twin} aligns the relevance metrics used in retrieval and downstream attention. TransAct V2~\cite{xia2025transact} strengthens interactions among retrieved behaviors, while MUSE~\cite{wu2025muse} incorporates multimodal semantics. Although preserving item-level access, these methods search the long history for each candidate, leaving online cost dependent on sequence length.

\textbf{Compression-based methods.} These methods summarize ultra-long histories into compact representations to bound candidate-dependent computation. C-Former~\cite{wang2025transformers} and DV365~\cite{lyu2025dv365} learn compact interest states or progressively reduced representations, while SequenceO1~\cite{guan2026sequenceo1} supports histories of up to 100K events through a fixed-size user-only sketch, cache reuse, and recent-history modeling. Although efficient, such compression may limit item-level access and discard sparse target-relevant evidence when the representation is constructed before the target is observed.

\textbf{Hybrid retrieval-and-compression methods.} These methods combine search and compression to balance efficiency and effectiveness. DILN~\cite{cai2025interest} first selects interactions by action type and then compresses the resulting subsequences with CNN-based encoders. In contrast, TWIN V2~\cite{si2024twin} compresses lifelong behaviors into cluster-structured representations before cluster-aware target retrieval. ADFM~\cite{li2022adversarial} pre-selects and stores top-$k$ behaviors for reuse during inference. While reducing online computation, these methods may limit item-level access or omit target-relevant evidence when their offline selection or compression criteria differ from downstream relevance.

\section{Methodology}\label{sec:method}

\providecommand{\citehint}[1]{\textcolor{red}{\scriptsize\,[#1]}}
\providecommand{\tofill}[1]{\textcolor{red}{\textbf{#1}}}

\subsection{Problem Formulation}
\label{sec:problem}

Let $v$ denote a target item, and let $\mathcal{H}=((x_j,\tau_j))_{j=1}^{L}$ denote the user's chronologically ordered lifetime behavior sequence, where $L$ is the sequence length, $x_j$ contains the item identity and side features of the $j$-th historical event, and $\tau_j$ is its timestamp. Repeated interactions with the same item occupy distinct sequence positions. Given the user and request context $c$, target features $x_v$, and history $\mathcal{H}$, the ranker estimates the probability of each behavior label $y^{(t)}$ for task $t$. We focus on ultra-long histories with lengths on the order of $10^5$, as considered in recent work~\cite{guan2026sequenceo1}.

Throughout, $C$ denotes the number of target items scored per request, $K$ the number of original event records retained by the user-level selector, and $M$ the number of records passed to final interest aggregation, where $M \ll K \ll L$.

\subsection{Framework Overview}
\label{sec:overview}

\modelname{} is an end-to-end framework that connects real-time sequence selection with target-aware ranking through two nested budgets (Figure~\ref{fig:architecture}). In Figure~\ref{fig:architecture}(a), during real-time training, a \textbf{Distilled Selector (DS)} scores each user's lifetime history using user features independent of the candidate item and writes $K$ retained event records to an intermediate store. The ranking graph in Figure~\ref{fig:architecture}(b) reads these records during both training and online inference. A \textbf{Target-aware Fine-grained Retriever (TFR)} selects $M$ records for each target using TWIN-style retrieval~\cite{chang2023twin}, and \textbf{Candidate-aware Interest Aggregation} summarizes them for the multi-task ranker. Online requests only read the retained history; they do not execute DS or rescan the full sequence. A typical configuration in our system uses $K=10{,}000$ and $M=100$.

\subsection{User-Level Sequence Compression}
\label{sec:compression}

DS reduces the searchable history from $L$ to $K$ before target-aware ranking; the empirical basis for compressing $10^5$-level histories to $10^4$ retained records is examined in Sec.~\ref{sec:exp_diagnostics}. Its input $a$ contains only user features, such as user identity embeddings, and does not depend on the candidate item. The selection is user-only, so the stored records are reused across requests and candidates.

\subsubsection{Selection}

DS embeds each historical event from its features and timestamp as $z^S_j=f^S_{\mathrm{hist}}(x_j,\tau_j)$ and the user from profile features $a$ as $e=f_{\mathrm{user}}(a)$.
The score is the mean inner product over $H_s$ heads:
\vspace{-2pt}
\begin{equation}
    s^S_j=\frac{1}{H_s}\sum_{h=1}^{H_s}
    \bigl\langle W^S_{Q,h}e,\;W^S_{K,h}z^S_j\bigr\rangle.
    \label{eq:student_score}
\end{equation}
\vspace{-2pt}
All equations range over real events; padded positions in batched tensors are masked out of selection and normalization. DS retains the highest-scoring positions,
\vspace{-2pt}
\begin{equation}
    \mathcal{I}^{K}=\operatorname*{Top\text{-}K}_{j\in[1,L]}(s^S_j),
    \qquad
    \widetilde{\mathcal{H}}=
    \bigl((x_j,\tau_j)\bigr)_{j\in\mathcal{I}^{K}},
    \label{eq:cas_topk}
\end{equation}
\vspace{-2pt}
where $\mathcal{I}^{K}$ is a score-ordered list of event positions. Unlike latent-token compression, which represents history through mixtures of features, this operation retains original records with their item IDs, timestamps, and side features intact. They remain individually addressable for downstream target-to-item matching, including evidence for long-tail targets and dispersed interests.

\subsubsection{Target-aware distillation}
\label{sec:distillation}

A selector conditioned only on user features cannot adapt to the target, yet relevance in ranking is target-dependent. We bridge this gap with a candidate-aware teacher trained over the full history in the selector graph. The teacher uses its own event encoder $z^T_j=f^T_{\mathrm{hist}}(x_j,\tau_j)$, with embeddings and transformations independent of the student's. For each training target $v$, it averages relevance logits over $H_t$ heads:
\vspace{-2pt}
\begin{equation}
    s^T_{v,j}=\frac{1}{H_t}\sum_{h=1}^{H_t}
    \bigl\langle W^T_{Q,h}f_T(c,x_v),\;W^T_{K,h}z^T_j\bigr\rangle,
    \label{eq:teacher_score}
\end{equation}
\vspace{-2pt}
where $f_T$ encodes context and target features. Distillation aligns the student's selection with the teacher's per-target relevance.

\paragraph{Distribution matching.} For a batch $\mathcal B$, let $b$ index a user--target sample with user $u_b$, target $v_b$, and history length $L_b$. Write $s^T_{b,j}$ for Eq.~\eqref{eq:teacher_score} evaluated at $v_b$, and $s^S_{b,j}$ for the user-only student score of $u_b$, which Eq.~\eqref{eq:student_score} shares across all samples of that user. With temperature $T$, we define $p^T_{b,j}=\operatorname{softmax}_{j}(s^T_{b,j}/T)$ and $p^S_{b,j}=\operatorname{softmax}_{j}(s^S_{b,j}/T)$. These normalized distributions correspond to the teacher and student attention weights in Figure~\ref{fig:architecture}(a), rather than the raw relevance logits. The distillation loss is
\vspace{-2pt}
\begin{equation}
    \mathcal L_{\mathrm{KD}}=
    -\frac{T^2}{|\mathcal B|}\sum_{b\in\mathcal B}\sum_{j=1}^{L_b}
    \operatorname{sg}[p^T_{b,j}]\log p^S_{b,j},
    \label{eq:kd_loss}
\end{equation}
\vspace{-2pt}
where $\operatorname{sg}$ denotes stop-gradient. Rather than using only a binary indication of whether each record belongs to the teacher's Top-$K$, the soft targets encode how strongly the teacher prefers each position in the history.

\paragraph{Boundary-targeted auxiliary term.} Matching the full distribution can still leave teacher-preferred records outside the student's Top-$K$. We therefore add a ranking objective across the selection boundary: it raises the student scores of missed records preferred by the teacher relative to those of less relevant retained records, encouraging their inclusion in the selected history.

For each sample $b$, form a sharper teacher distribution $\tilde p^T_b=\operatorname{softmax}(s^T_b/T_m)$, with $T_m<T$. Against the user's student-selected set $\mathcal I^K_{u_b}$, mine up to $n$ positions in each region:
\vspace{-2pt}
\begin{equation}
\begin{aligned}
    (j^+_{b,1},\ldots)
    &=\operatorname*{Top\text{-}n}_{j\notin\mathcal I^K_{u_b}}
        (\tilde p^T_{b,j}),\\
    (j^-_{b,1},\ldots)
    &=\operatorname*{Top\text{-}n}_{j\in\mathcal I^K_{u_b}}
        (-\tilde p^T_{b,j}).
\end{aligned}
    \label{eq:hard_pairs}
\end{equation}
\vspace{-2pt}
Missed positions are ordered by decreasing teacher mass and selected positions by increasing teacher mass. Rank-aligned pairs form the set $\mathcal V_b$; for $i\in\mathcal V_b$, the weights are
\vspace{-2pt}
\begin{equation}
    \omega_{b,i}=\operatorname{sg}\!\left[
      \max\!\left(\tilde p^T_{b,j^+_{b,i}}-
                        \tilde p^T_{b,j^-_{b,i}},0\right)\right],
    \quad i\in\mathcal V_b.
    \label{eq:pair_weight}
\end{equation}
\vspace{-2pt}
The batch-normalized loss is
\vspace{-2pt}
\begin{equation}
    \mathcal L_{\mathrm{aux}}=
    \frac{\displaystyle
      \sum_{b\in\mathcal B}\sum_{i\in\mathcal V_b}\omega_{b,i}
      \operatorname{softplus}\!\left(
      \frac{s^S_{b,j^-_{b,i}}-s^S_{b,j^+_{b,i}}+\delta}{T_{\mathrm{rank}}}\right)
    }{\displaystyle
      \sum_{b\in\mathcal B}\sum_{i\in\mathcal V_b}\omega_{b,i}+\epsilon},
    \label{eq:aux_loss}
\end{equation}
\vspace{-2pt}
where the margin $\delta$ encourages a missed positive's student score to exceed its paired negative's score by $\delta$, $T_{\mathrm{rank}}>0$ controls the ranking scale, and $\epsilon>0$ stabilizes the denominator. Invalid pairs are excluded through $\mathcal V_b$, and weights vanish whenever the teacher does not prefer the missed position. Mining and weights are detached, while gathered student logits remain differentiable. Thus each target contributes its own pairs without propagating gradients through discrete selection or into the teacher.

\subsection{Target-Aware Retrieval and Aggregation}
\label{sec:retrieval}

\begin{algorithm}[!t]
\caption{Real-time training of \modelname{}}
\label{alg:training}
\small
\begin{algorithmic}[1]
\Require user--target training stream; histories $\mathcal H$; budgets $K,M$; weights $\alpha,\beta,\gamma$
\Ensure DS; relevance parameters $\Theta_{\mathrm{rel}}$; multi-task ranker
\Statex \textbf{Selector graph} (concurrent with the ranking graph)
\For{each batch $\mathcal B$, grouped by user}
  \State Encode each history independently as $z^S_j$ and $z^T_j$
  \State Compute user-only $s^S$ and $\mathcal I^K$ by Eqs.~\eqref{eq:student_score},~\eqref{eq:cas_topk}
  \State Form student distribution $p^S$ at temperature $T$
  \For{each user--target sample $b\in\mathcal B$}
    \State Compute head-mean teacher logits by Eq.~\eqref{eq:teacher_score}
    \State Form detached $p^T_b$ at $T$ and $\tilde p^T_b$ at $T_m$
    \State Mine highest-mass missed and lowest-mass
    \Statex \hspace{\algorithmicindent}\hspace{\algorithmicindent}selected positions using $\tilde p^T_b$ by Eq.~\eqref{eq:hard_pairs}
    \State Rank-align pairs; form detached gap weights
    \Statex \hspace{\algorithmicindent}\hspace{\algorithmicindent}$\omega_{b,i}$ by Eq.~\eqref{eq:pair_weight}
  \EndFor
  \State Compute $\mathcal L_{\mathrm{KD}}$ over all valid history positions
  \State Compute $\mathcal L_{\mathrm{aux}}$ from weighted student-logit
  \Statex \hspace{\algorithmicindent}differences by Eq.~\eqref{eq:aux_loss}
  \State Update teacher by $\alpha\mathcal L_{\mathrm{pred}}$
  \State Update DS by $\beta\mathcal L_{\mathrm{KD}}+\gamma\mathcal L_{\mathrm{aux}}$
  \State Write retained records $\widetilde{\mathcal H}$ to the shared store
  \State Synchronize teacher $\Theta_{\mathrm{rel}}$ to TFR and aggregation
\EndFor
\Statex \textbf{Ranking graph} (concurrent with the selector graph)
\For{each batch of user--target samples}
  \State Read $\widetilde{\mathcal H}$ from the shared store for each user
  \For{each target $v$}
    \State Retrieve $\mathcal R^M_v$ using $\Theta_{\mathrm{rel}}$ (Sec.~\ref{sec:retrieval})
    \State Aggregate with target attention using $\Theta_{\mathrm{rel}}$
    \State Predict behavior probabilities with the ranker
  \EndFor
  \State Train the ranking graph with $\mathcal L_{\mathrm{rank}}$
\EndFor
\end{algorithmic}
\end{algorithm}

TWIN~\cite{chang2023twin} uses consistent relevance-scoring parameters in its general and exact search units (GSU/ESU). We follow this strategy and extend parameter reuse to the full-history teacher: its learned query/key relevance parameters, denoted by $\Theta_{\mathrm{rel}}$, are exported to both TFR and the target-attention scorer.

Given $\widetilde{\mathcal H}$ and target $v$, TFR selects $M$ records by their relevance scores using TWIN-style multi-head retrieval. Denote their positions by $\mathcal R^M_v$. The retrieved records are then aggregated with TWIN-style target attention and fed into the final ranker.

\subsection{Training Objectives}
\label{sec:objective}

The selector graph trains the full-history teacher and DS with
\vspace{-2pt}
\begin{equation}
    \mathcal L_{\mathrm{sel}}
    =\alpha\mathcal L_{\mathrm{pred}}
     +\beta\mathcal L_{\mathrm{KD}}
     +\gamma\mathcal L_{\mathrm{aux}},
    \label{eq:stage1_loss}
\end{equation}
\vspace{-2pt}
where $\mathcal L_{\mathrm{pred}}$ is the teacher's multi-task prediction loss over behavior labels. The objective covers the whole selector graph: only $\mathcal L_{\mathrm{pred}}$ updates the teacher, while KD and auxiliary supervision train DS. Independent embeddings and detached teacher targets prevent student losses from updating the teacher. The teacher is therefore a live prediction branch rather than a separately pretrained, frozen model.\looseness=-1

The ranking graph uses its own multi-task prediction objective $\mathcal L_{\mathrm{rank}}$, with no additional distillation loss. Both graphs train on the same real-time dataset and share the intermediate store: the selector graph writes retained records and the ranking graph reads them. Algorithm~\ref{alg:training} summarizes the two concurrent training processes.

\subsection{Complexity Analysis}
\label{sec:complexity}

We separate training from online inference and count the main sequence-interaction operations, excluding feature projections, discrete selection/mining, store access, and the rest of the ranker. Let $H_s,H_t,H_r,H_a$ be the head counts of DS, teacher, retrieval, and aggregation. The corresponding widths $d_s,d_t,d_r,d_a$ reflect the operators actually executed; in particular, $d_r$ is the deployed per-head scoring width, and $d_a$ the per-head width used inside aggregation.

\paragraph{Training.} For a group of $B$ targets sharing one user's length-$L$ history, DS scoring costs $\mathcal O(LH_sd_s)$, while full-history teacher interactions cost $\mathcal O(BLH_td_t)$, with $d_t$ covering teacher scoring and value aggregation. Distribution matching adds $\mathcal O(BL)$ and the loss over at most $n$ mined pairs adds $\mathcal O(Bn)$; these do not include pair mining itself. The ranking graph operates on the retained $K$ records. Thus training still includes length-dependent work, even though user-only computation is shared.

\paragraph{Online inference.} A request reads the stored $K$ records and evaluates $C$ targets. Its principal sequence-interaction cost is
\vspace{-2pt}
\begin{equation}
    \Delta_{\text{\modelname{}}}
    =\mathcal O(C H_r K d_r)+\mathcal O(C H_a M d_a).
    \label{eq:cost}
\end{equation}
\vspace{-2pt}
The shared record read transfers $\mathcal O(K)$ data at fixed record width. Neither DS nor the full-history teacher executes online, so for fixed $K$ and $M$ the online branch is independent of the original history length $L$. In contrast, applying direct target-aware retrieval, as in TWIN~\cite{chang2023twin}, to the full length-$L$ history incurs an interaction cost linear in both $C$ and $L$.

\subsection{Deployment Optimizations}
\label{sec:deployment}

Beyond the sequence budgets, deployment must control repeated feature processing, storage, and data movement. We use three \mbox{complementary} optimizations.

\subsubsection{Common-feature sharing}

We group multiple training samples from the same user within a batch and load their common user-side features, including the long behavior history, into GPU memory only once per user group. All samples in the group share this in-memory copy instead of materializing a separate copy for each user--target sample. This reduces the GPU memory occupied by duplicated common features, which is particularly costly for ultra-long histories. Target-specific features and computations remain sample-specific. This organization follows the user-level batching principle used by recent ultra-long-sequence systems~\cite{guan2026make,guan2026sequenceo1}.

\subsubsection{Compact record encoding}

The record schema depends on the application. In our production example, eight fields represented as int64 values require $8\times8=64$ bytes per event, or 640~KB at $K=10{,}000$. Rather than reserving 64 bits for every field, we assign each field a fixed storage width sufficient for its value range and pack the fields consecutively into a record. The field widths are fixed by the application schema, not chosen separately for each event. In our deployed schema, the packed fields total 43 bytes per event, giving approximately 0.43~MB of record payload per user, about a third less than the int64 layout. The sequence is stored as one contiguous user-keyed value, enabling a single logical store operation. At a scale of one billion users, each with $10{,}000$ stored records, this encoding would reduce single-copy record payload from 640~TB to 430~TB, saving approximately 210~TB before replication and storage-system overhead.\looseness=-1

\subsubsection{Key-side projection absorption}

Computing target--history relevance conventionally projects both query and history vectors. Since each history position requires a key projection, repeating this work for every request introduces sequence-length-dependent computation on the online path. Caching projected keys avoids repeated projection but requires a cache whose size also grows with the number of history positions. We instead transfer the key projection to the query side through the identity
\vspace{-2pt}
\begin{equation}
    \langle qW_Q,\;hW_K\rangle
    =\langle (qW_Q)W_K^\top,\;h\rangle.
    \label{eq:absorption}
\end{equation}
\vspace{-2pt}
This avoids explicitly projecting each retained history vector into a key and materializing the resulting per-position tensor, eliminating the need to maintain a projected-key cache.

\section{Experiments}

\begin{table*}[t]
 \centering
 \def\TableOneScale{0.90}
 \caption{Performance comparison among different methods on KuaiRand-1K and Industrial Dataset.}
 \label{tab:overall_main}
 \setlength{\tabcolsep}{5pt}
 \normalsize
 \renewcommand{\arraystretch}{1.14}
 \resizebox{\TableOneScale\textwidth}{!}{%
 \begin{tabular*}{\textwidth}{@{\extracolsep{\fill}}lrrrrrrrr@{}}
  \toprule
  & \multicolumn{4}{c}{KuaiRand-1K} & \multicolumn{4}{c}{Industrial Dataset} \\
  \cmidrule(lr){2-5}\cmidrule(lr){6-9}
  Method & Click & Like & Comment & Long-view & Effect-view & Like & Comment & Long-view \\
  \midrule
DIN~\cite{zhou2018deep} & 0.6574 & 0.6479 & 0.6004 & 0.6917 & 0.7050 & 0.7334 & 0.7738 & 0.7498 \\
SIM~\cite{pi2020search} & 0.6592 & 0.6585 & 0.6034 & 0.6929 & 0.7055 & 0.7383 & 0.7791 & 0.7522 \\
TWIN~\cite{chang2023twin} & 0.6607 & 0.6594 & 0.6185 & 0.6943 & 0.7057 & 0.7399 & 0.7805 & 0.7511 \\
STCA~\cite{guan2026make} & 0.6612 & \underline{0.6617} & 0.6226 & 0.6944 & 0.7069 & 0.7388 & 0.7822 & 0.7540 \\
TWIN V2~\cite{si2024twin} & 0.6611 & 0.6613 & 0.6243 & \underline{0.6945} & 0.7070 & \underline{0.7402} & \underline{0.7824} & 0.7527 \\
SequenceO1~\cite{guan2026sequenceo1} & \underline{0.6615} & 0.6607 & \underline{0.6252} & 0.6942 & \underline{0.7077} & 0.7389 & 0.7818 & \underline{0.7544} \\
\modelname{} & \textbf{0.6630} & \textbf{0.6656} & \textbf{0.6347} & \textbf{0.6967} & \textbf{0.7085} & \textbf{0.7424} & \textbf{0.7851} & \textbf{0.7553} \\
\midrule
Impro.  & +0.227\% & +0.589\% & +1.520\% & +0.317\% & +0.113\% & +0.297\% & +0.345\% & +0.119\% \\
  \bottomrule
 \end{tabular*}%
 }
 \\
\small{Here, the best and second best results are marked in \textbf{bold} and \underline{underlined}, respectively.}
\end{table*}


\subsection{Experimental Settings}
In this section, we evaluate the performance of ECHO using public and industrial datasets through a series of experiments, mainly to answer four research questions as follows:
\begin{itemize}
 \item \textbf{RQ1:} How does ECHO perform compared to state-of-the-art methods on real-world datasets?
 \item \textbf{RQ2:} How effectively does ECHO select and preserve relevant behaviors from ultra-long user histories?
 \item \textbf{RQ3:} How efficiently does ECHO scale to longer user behavior sequences?
 \item \textbf{RQ4:} How do different components and hyperparameters affect ECHO's performance?
\end{itemize}

\label{sec:exp_setup}
\paragraph{Datasets and metrics.}
We evaluate \modelname{} on the KuaiRand-1K benchmark~\cite{gao2022kuairand} and Kuaishou’s industrial
short-video recommendation platform, which is the largest recommendation scenario in Kuaishou and serves more than \textbf{400 million
daily active users}, generating over \textbf{50 billion interaction logs
per day}. For the KuaiRand-1K benchmark, we evaluate on four tasks: \textit{Click}, \textit{Long-view}, \textit{Like}, and \textit{Comment}. For the Kuaishou industrial dataset, we follow practical deployment requirements and focus on four key prediction tasks: \textit{Effective-view}, \textit{Long-view}, \textit{Like}, and \textit{Comment}. We adopt the \textbf{GAUC}~\cite{zhou2018deep,zhang2026smes} to evaluate the performance of our model, which is the most critical offline metric in our short-video service. Notably, given the stability provided by the massive
scale of training samples, a relative GAUC improvement of 0.05\% is regarded as highly significant and can bring substantial business
impact~\cite{zhang2026smes}.

\paragraph{Baselines.}
We compare six representative sequence models: DIN~\cite{zhou2018deep}, SIM~\cite{pi2020search}, TWIN~\cite{chang2023twin}, STCA~\cite{guan2026make}, TWIN V2~\cite{si2024twin}, and
SequenceO1~\cite{guan2026sequenceo1}. More details about baseline descriptions and implementation are provided in Appendix~\ref{app:baselines} and Appendix~\ref{app:implementation}, respectively.

\subsection{Overall Performance (RQ1)}
\label{sec:exp_main}

Table~\ref{tab:overall_main} compares ECHO with six representative sequence modeling methods on KuaiRand-1K and the industrial dataset. ECHO consistently achieves the best GAUC across all eight prediction tasks, with relative improvements of 0.227\%–1.520\% on KuaiRand-1K and 0.113\%–0.345\% on the industrial dataset over the strongest baseline for each task. These results demonstrate the effectiveness of ECHO across different datasets and feedback objectives.

Among the baselines, search-based methods such as SIM and TWIN generally outperform DIN, which models only recent user behaviors, highlighting the importance of incorporating longer interaction histories for accurate user interest modeling. More advanced methods, including STCA, TWIN V2, and SequenceO1, achieve further improvements on several tasks through end-to-end long-sequence modeling or history compression. These results suggest that effectively exploiting historical information under constrained computational budgets is critical to improving recommendation performance as user behavior sequences grow longer.

ECHO further improves upon these methods across all tasks. In particular, on KuaiRand-1K, ECHO increases the Comment GAUC from 0.6252 to 0.6347, achieving a relative improvement of 1.520\%. On the industrial dataset, the corresponding GAUC increases from 0.7824 to 0.7851, representing a 0.345\% improvement. Unlike methods that truncate historical sequences or compress historical behaviors into aggregated representations, ECHO learns a user-level behavior selection strategy through target-aware distillation. This design reduces the online retrieval space while preserving original behavior records for fine-grained target matching. Furthermore, the shared relevance metric aligns the subsequent retrieval and interest aggregation stages, enabling ECHO to exploit prediction-relevant historical evidence within a bounded computational budget.

The improvements are also consistent across different feedback objectives. On KuaiRand-1K, ECHO achieves relative gains of 0.589\% and 1.520\% on Like and Comment, respectively. Similar improvements are observed on the industrial dataset, with gains of 0.297\% and 0.345\% on the corresponding tasks. ECHO also improves viewing-related predictions, achieving relative gains of 0.113\% on Effective-view and 0.119\% on Long-view in the industrial dataset. These results demonstrate that the benefits of ECHO extend beyond specific interaction signals, supporting its effectiveness across diverse recommendation objectives.

\subsection{Selected History Analysis (RQ2)}
\label{sec:exp_budgets}
\label{sec:exp_diagnostics}

\begin{figure}[t]
 \centering
 \scalebox{\FigThreeScale}{\includegraphics[width=0.9\linewidth]{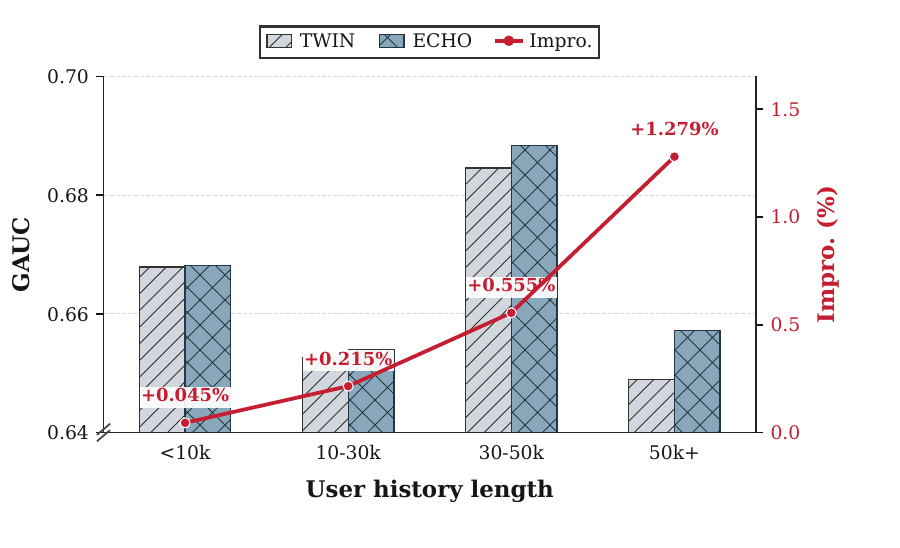}}
 \caption{Click GAUC of TWIN and ECHO across different user history lengths on KuaiRand-1K. Relative improvements are shown by the line.}
 \Description{TWIN and ECHO GAUC bars by user history length, with relative GAUC improvement over TWIN shown on the right axis. The GAUC axis ranges from 0.64 to 0.70.}
 \label{fig:history_length}
\end{figure}

To understand how ECHO benefits from ultra-long user histories, we investigate the effectiveness of the Distilled Selector (DS) from three perspectives: downstream prediction performance, preservation of relevant historical behaviors, and sensitivity to the selection budget. We compare DS with recency-based truncation and uniform random selection, using the target-aware teacher as a reference for evaluating the quality of retained historical evidence.

We first examine how the benefit of learned selection varies with history length. As shown in Figure~\ref{fig:history_length}, ECHO consistently outperforms TWIN with a recent 10K history, and the improvement becomes more pronounced as the available history grows. Specifically, the relative GAUC gain increases from 0.045\% for histories shorter than 10K to 1.279\% for histories exceeding 50K. This trend suggests that a fixed recent-history window increasingly overlooks useful historical evidence as user behaviors accumulate, whereas ECHO can identify relevant interactions from a broader history. The advantage persists under smaller sequence budgets. With retention and retrieval budgets of 5K and 50, respectively, ECHO achieves a GAUC of 0.6618, compared with 0.6603 for recent-5K TWIN. Even when the budgets are reduced to 2K and 20, ECHO maintains an improvement from 0.6585 to 0.6599. These results indicate that the effectiveness of ECHO stems not only from accessing longer histories, but also from allocating the limited sequence budget to more informative behaviors.

\begin{figure}[t]
 \centering
 \scalebox{\FigFourScale}{\includegraphics[width=\linewidth]{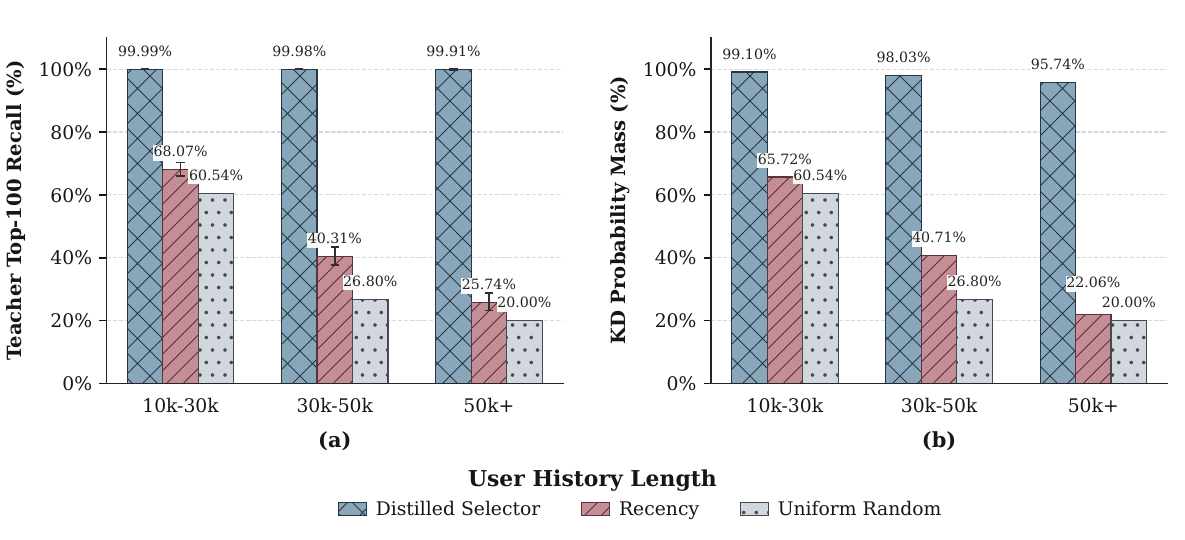}}
 \caption{Preservation of teacher-relevant behaviors by 10K retained records across user history lengths. (a) Recall of the teacher's Top-100 behaviors. (b) Fraction of the teacher's full-history KD probability mass preserved.}
 \Description{Two grouped bar charts compare the Distilled Selector, Recency, and Uniform Random. The left panel shows Teacher Top-100 Recall; the right panel shows retained KD probability mass. Both use the same history-length groups and a shared legend.}
 \label{fig:history_preservation}
\end{figure}

\begin{figure}[t]
 \centering
 \scalebox{\FigSixScale}{\includegraphics[width=0.8\linewidth]{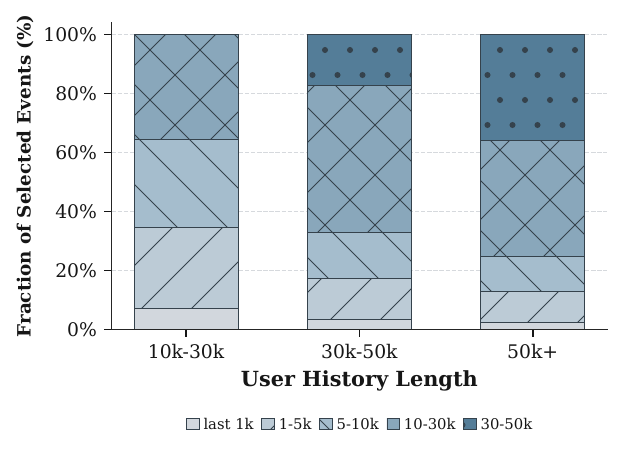}}
 \caption{Temporal distribution of behaviors retained by the distilled selector. Age is measured by the number of preceding interactions rather than elapsed time.}
 \Description{Stacked age distributions of 10000 selected events for three history-length groups.}
 \label{fig:event_age}
\end{figure}

We next investigate whether DS successfully preserves the historical behaviors favored by the target-aware teacher. Figure~\ref{fig:history_preservation}(a) reports the recall of the teacher's Top-100 relevant behaviors within the 10K retained events. DS achieves an average recall of 99.976\%, substantially outperforming recency-based and random selection, which achieve 57.905\% and 49.443\%, respectively. Notably, its recall remains above 99.9\% across all history-length groups, while recency-based selection deteriorates considerably as history length increases. Figure~\ref{fig:history_preservation}(b) further evaluates the proportion of the teacher's relevance distribution preserved by the selected events. DS retains 98.509\% of the total relevance mass on average, compared with only 55.913\% for recency-based selection, and maintains 95.74\% even for the longest histories. These results demonstrate that target-aware distillation enables the candidate-agnostic selector to preserve most of the historical evidence emphasized by the teacher, despite substantially reducing the searchable history.

\begin{figure}[t]
 \centering
 \scalebox{\FigFiveScale}{\includegraphics[width=\linewidth]{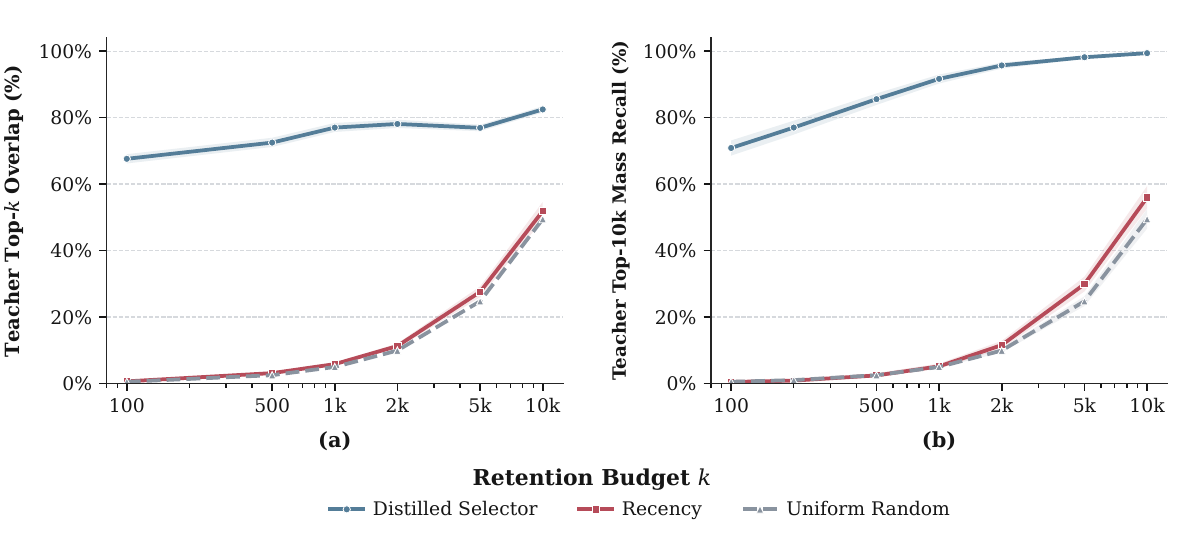}}
 \caption{Selection quality under different retention budgets. (a) Overlap between selector and teacher Top-$k$ sets at equal budgets. (b) Recall of the teacher's Top-10K attention mass.}
 \Description{Two line charts compare the Distilled Selector, Recency, and Uniform Random across retention budgets. The left panel shows same-budget Top-k overlap; the right panel shows Teacher Top-10K attention mass recall. Original confidence intervals and a shared legend are retained.}
 \label{fig:selection_budget}
\end{figure}

We further analyze the temporal distribution of selected behaviors and their agreement with the teacher. As illustrated in Figure~\ref{fig:event_age}, 46.08\% of the behaviors retained by DS fall outside the most recent 10K interactions, with this proportion increasing to 75.09\% for the longest-history group. This confirms that DS does not simply approximate recency-based truncation, but actively selects informative interactions from older portions of the history. Meanwhile, Figure~\ref{fig:selection_budget}(a) compares DS and the teacher under identical Top-\(k\) selection budgets. Their overlap reaches 67.58\% at \(k=100\) and 82.45\% at \(k=10K\). Together with the near-perfect recall of the teacher's Top-100 behaviors within the retained 10K events, these findings suggest that DS need not reproduce the teacher's exact fine-grained ranking. Instead, it can preserve a sufficiently broad set of relevant original behaviors, allowing the subsequent target-aware retriever to perform more precise matching for each candidate.

Finally, we investigate how effectively DS preserves attention mass under different retention budgets. As shown in Figure~\ref{fig:selection_budget}(b), using the teacher's Top-10K behaviors as the reference, DS captures 70.84\%, 91.65\%, and 99.38\% of the reference attention mass with retention budgets of 100, 1K, and 10K, respectively. In contrast, recency-based selection retains only 0.43\%, 5.14\%, and 55.93\% under the same budgets. The substantial gap indicates that learned selection can concentrate teacher-attended historical evidence into a much smaller subset of original behavior records. This property is particularly important for ECHO's cascaded design: the user-level selector reduces the searchable history while preserving high-relevance events, and the subsequent target-aware retriever performs fine-grained matching within the retained set. Overall, these results validate the effectiveness of distillation-guided selection in balancing historical evidence preservation and bounded online retrieval costs.

\subsection{Efficiency and Length Scaling (RQ3)}
\label{sec:exp_efficiency}

\begin{figure}[t]
 \centering
 \scalebox{\FigSevenScale}{\includegraphics[width=0.9\linewidth]{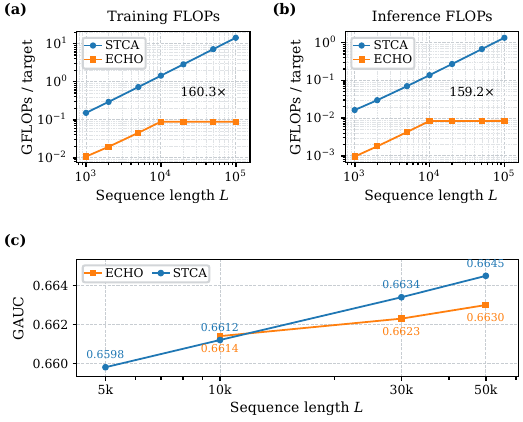}}
 \caption{History-length scaling. (a,b) Analytical training and inference sequence-branch FLOPs. (c) GAUC from a separate length study.}
 \Description{A single-column figure with two FLOPs panels above one wide GAUC panel, comparing ECHO and STCA.}
 \label{fig:efficiency_scaling}
\end{figure}

To evaluate the computational efficiency and scalability of ECHO for ultra-long user behavior modeling, we compare it with STCA, an end-to-end long-sequence modeling method, in terms of computational cost and predictive performance. STCA directly models historical behaviors through stacked target-aware attention layers, whereas ECHO adopts a cascaded architecture that combines user-level behavior selection with target-aware retrieval, restricting downstream computation to a bounded set of historical events. We estimate the training and inference FLOPs based on the sequence modeling architectures of both methods, with detailed calculations and configuration assumptions provided in Appendix~\ref{app:flops}.

Figures~\ref{fig:efficiency_scaling}(a) and (b) illustrate the computational costs of both methods as the history length increases. While the training and inference costs of STCA grow substantially with sequence length, ECHO exhibits much slower growth. This advantage stems from decoupling user-level behavior selection from candidate-dependent interest modeling, allowing subsequent retrieval and attention operations to process only a bounded number of historical events. At a history length of 100K, ECHO reduces the estimated sequence-branch FLOPs by approximately 160.3× and 159.2× during training and inference, respectively, compared with STCA. These results demonstrate the computational efficiency of the cascaded architecture when scaling to ultra-long behavior sequences.

We further investigate how predictive performance scales with history length. As shown in Figure~\ref{fig:efficiency_scaling}(c), both ECHO and STCA achieve consistent GAUC improvements as the available history length increases, indicating that longer histories provide additional information for user interest modeling. Specifically, the GAUC of ECHO increases from 0.6614 at 10K to 0.6630 at 50K, while STCA improves from 0.6598 at 5K to 0.6645 at 50K. Although STCA has a slight advantage at longer histories by directly modeling the full history, ECHO continues to benefit from longer sequences by preserving informative historical behaviors through its distilled selector and performing target-aware retrieval within a bounded sequence budget.

Overall, ECHO effectively reduces candidate-dependent sequence computation while retaining the predictive benefits of long-history modeling. By shifting history selection to a reusable user-level stage, ECHO enables efficient access to increasingly long behavior histories without requiring full-history target-aware interactions for every candidate. These results highlight its potential for scalable ultra-long sequence modeling under the computational constraints of industrial recommendation systems.

\subsection{Ablation Study (RQ4)}
\label{sec:exp_temperature}

We evaluate ECHO by ablating its training objectives, distillation temperature, and downstream ranking architectures.

To examine the contribution of distillation supervision, we remove both the knowledge distillation loss \(\mathcal{L}_{KD}\) and the boundary-targeted auxiliary loss \(\mathcal{L}_{aux}\), equip the student with an independent prediction head, and train it using prediction loss alone. As shown in Table~\ref{tab:loss_ablation}, the full ECHO model consistently outperforms the prediction-only variant across all four tasks on KuaiRand-1K, achieving relative GAUC improvements of 0.272\%, 0.925\%, 2.371\%, and 0.317\% on Click, Like, Comment, and Long-view, respectively. The largest improvement is observed on Comment, where GAUC increases from 0.6200 to 0.6347. These results demonstrate the importance of distillation-based supervision for effective history selection. By transferring target-aware relevance knowledge from the teacher to the user-level selector, the distillation objective guides the preservation of informative historical behaviors, while the boundary-targeted auxiliary objective further encourages teacher-preferred events to be included in the retained set. Together, these objectives improve history selection and downstream predictions.

Furthermore, we investigate the sensitivity of ECHO to the distillation temperature and evaluate its applicability across different downstream architectures. Due to space limitations, detailed experimental results are provided in Appendix~\ref{app:additional_results}.

\begin{table}[t]
 \centering
 \def\LossAblationTableScale{1.0} 
 \caption{Training-objective ablation on KuaiRand-1K (GAUC). The prediction-only variant removes both KD and auxiliary losses. Improvements are relative to this variant.}
 \label{tab:loss_ablation}
 \small
 \setlength{\tabcolsep}{3pt}
 \renewcommand{\arraystretch}{1.15}
 \resizebox{\LossAblationTableScale\linewidth}{!}{%
 \begin{tabular}{@{}lrrrr@{}}
  \toprule
  Method & Click & Like & Comment & Long-view \\
  \midrule
  ECHO (full) & \textbf{0.6630} & \textbf{0.6656} & \textbf{0.6347} & \textbf{0.6967} \\
  Prediction loss only & 0.6612 & 0.6595 & 0.6200 & 0.6945 \\
  \midrule
  Impro. & +0.272\% & +0.925\% & +2.371\% & +0.317\% \\
  \bottomrule
 \end{tabular}%
 }
\end{table}

\subsection{Online A/B Testing}
\label{sec:exp_online}

To assess the practical effectiveness of ECHO, we conduct a seven-day online A/B test in the Kuaishou short-video recommendation scenario described above, with 5\% of production traffic assigned to the treatment bucket.

Notably, the existing production system already incorporates advanced long-sequence modeling techniques, including TWIN-based behavior retrieval and sequence compression mechanisms similar to C-Former and SequenceO1. Rather than replacing a simple short-history baseline, ECHO is integrated into this established recommendation system to further enhance the utilization of ultra-long user histories. During online inference, the ranking system reads the retained historical records and performs target-aware retrieval and interest aggregation, while the user-level selector is maintained by the real-time training pipeline. The deployment is further supported by shared user inputs, compact record encoding, and key-side projection absorption, as detailed in Section~\ref{sec:deployment}.
Compared with the existing production systems, ECHO achieved \textbf{+0.240\%} and \textbf{+0.373\%} increases in \textit{Total App Usage Time} on Kuaishou and Kuaishou Lite, respectively, demonstrating its benefit to user engagement in production environments.

\section{Conclusion}

In this paper, we propose ECHO, a distillation-guided framework for ultra-long user behavior sequence modeling in industrial recommendation systems. ECHO employs a user-level Distilled Selector to identify and retain informative original behavior records from user histories of up to 100K events, followed by a Target-aware Fine-grained Retriever for downstream interest modeling. Through target-aware distillation and relevance parameter sharing, ECHO reduces candidate-dependent computation while preserving fine-grained matching capabilities over historical behaviors.
We conducted extensive experiments on KuaiRand-1K and an industrial dataset. ECHO achieves the best GAUC across all eight prediction tasks. Further analyses demonstrate the effectiveness of the distilled selector in preserving highly relevant historical behaviors, as well as the computational efficiency and scalability of ECHO for ultra-long sequence modeling. Moreover, we conducted seven-day online A/B tests in two production scenarios, Kuaishou and Kuaishou Lite. Compared with the existing production systems, ECHO achieved \textbf{+0.240\%} and \textbf{+0.373\%} increases in \textit{Total App Usage Time}, respectively, demonstrating the practical effectiveness of ECHO in large-scale recommendation systems.

\balance

\newpage
\bibliographystyle{ACM-Reference-Format}
\bibliography{reference}

\clearpage
\nobalance 
\appendix
\section{Additional Experimental Details}
\label{app:details}

\subsection{Baselines}
\label{app:baselines}
To comprehensively evaluate \modelname{}, we compare it with six representative user behavior sequence modeling methods, categorized according to how they process historical interactions.

(1) \textit{Short-history modeling.}
DIN~\cite{zhou2018deep} employs target-aware attention to capture user interests from recent interactions. Due to the computational cost of attention over long sequences, it typically models a limited window of recent behaviors.

(2) \textit{Retrieval-based long-sequence modeling.}
SIM~\cite{pi2020search} introduces a two-stage framework consisting of a General Search Unit (GSU) and an Exact Search Unit (ESU), which first retrieves target-relevant behaviors from long-term histories and then performs fine-grained interest modeling. TWIN~\cite{chang2023twin} further aligns the relevance metrics between the GSU and ESU, addressing the inconsistency between retrieval and downstream interest modeling.

(3) \textit{End-to-end long-sequence modeling.}
STCA~\cite{guan2026make} employs stacked target-aware cross-attention layers, together with batching optimizations and sequence-length extrapolation, to efficiently model user behavior histories at the 10K scale.

(4) \textit{Compression-based ultra-long sequence modeling.}
TWIN V2~\cite{si2024twin} compresses lifelong user behaviors into cluster-structured representations and performs cluster-aware target attention to extend the range of historical interactions available for modeling. SequenceO1~\cite{guan2026sequenceo1} constructs a fixed-size user history sketch and combines low-rank caching with recent-history modeling, enabling efficient modeling of user behavior sequences at the 100K scale while bounding candidate-dependent computation.

\subsection{Implementation Details}
\label{app:implementation}
In offline experiments, we configure the history length of each method according to its modeling characteristics. For DIN, the maximum history length is set to 100. For SIM and TWIN, the input history length of the General Search Unit (GSU) is limited to 10K behaviors. STCA is trained with an average sequence length of 2K and extrapolates to 10K during inference. TWIN V2 clusters the most recent 50K behaviors into approximately 10K representative units. SequenceO1 compresses the same behaviors into a fixed-size sketch of 100 tokens and uses cache reuse to support ultra-long sequence modeling. For \modelname{}, the maximum raw history length is set to 50K, from which the Distilled Selector (DS) retains 10K original behavior records for downstream target-aware retrieval. For all two-stage retrieval methods, the final retrieval budget is set to 100 behaviors for subsequent interest modeling.

In industrial online deployment, we further extend the maximum raw history length of \modelname{} to 100K behaviors to accommodate ultra-long user histories in real-world recommendation scenarios. The DS retains 10K original behavior records, while the Target-aware Fine-grained Retriever (TFR) retrieves 100 relevant behaviors for each candidate item. This configuration expands the accessible history range while keeping the downstream retrieval and interest modeling budgets fixed.

All experiments are conducted on NVIDIA A800 GPUs using a hybrid distributed training framework. For KuaiRand-1K, we adopt the Adam optimizer with a learning rate of $10^{-3}$ and a batch size of 256. For the industrial dataset, both sparse embedding parameters and dense network parameters are optimized using Adam, with learning rates of $0.001$ and $0.0005$, respectively, and a batch size of 3,000.

\subsection{Computational Cost Analysis}
\label{app:flops}
We provide the analytical FLOPs estimates used in
Fig.~\ref{fig:efficiency_scaling}(a) and (b).
We count dense matrix operations, with one multiply--accumulate
equal to two FLOPs.
Let $L$ denote the history length, $K=\min(L,10{,}000)$
the retention budget, and $R$ the number of candidates
sharing the same user history.
Candidate-independent computations are amortized over $R$,
while target-dependent operations are counted per candidate.
We set $R_{\mathrm{tr}}=40$ and $R_{\mathrm{inf}}=300$,
with assumed selection-cache hit rates
$h_{\mathrm{tr}}=0.5$ and $h_{\mathrm{inf}}=0.6$.
Cached selections are reused only for unchanged selector
versions and history snapshots, while trainable attention
activations are recomputed at each training step.

\paragraph{STCA}
We reconstruct the single-query attention reordering
and fusion architecture of STCA~\cite{guan2026make}
with $m=4$ layers, hidden width $d=128$,
$H=4$ heads, and SwiGLU expansion ratio $r=4$.
Each SwiGLU block costs $F=6rd^2$ FLOPs, yielding
$mFL/R$ per target for historical feed-forward networks.
Under reordered attention, each head performs scoring and
weighted reduction over the original $d$-dimensional
history, contributing $4mHLd$ FLOPs.
The remaining query-side projections and fusion operations
contribute
\begin{equation}
\begin{aligned}
B ={}& 8md^2+F+
\sum_{i=1}^{m-1}[2(i+1)d^2+F]\\
&+2(m+1)d^2+F\\
={}&(m+1)F+(m^2+11m)d^2.
\end{aligned}
\end{equation}
The resulting forward FLOPs per target are
\begin{equation}
\begin{aligned}
C_S(L,R)
&=\frac{mFL}{R}+4mHLd+B\\
&=\frac{1{,}572{,}864L}{R}
+8{,}192L+2{,}949{,}120.
\end{aligned}
\end{equation}
Both training and inference estimates use the full
history length $L$, rather than the shorter average
training length used for STCA's length extrapolation.

\paragraph{ECHO}
The counted ECHO computation consists of amortized
selector refresh and retained-history attention.
The frozen selector uses a $624\rightarrow128$ input
projection and grouped query-side transformations
of $128\rightarrow256\rightarrow185$,
where the final dimension includes a bias coordinate.
The input projection precedes user grouping, while the
subsequent transformations and history scan are shared.
With linear history-side projections absorbed into
the query side, the expected refresh cost is
\begin{equation}
S(L,R,h)=(1-h)\left[
159{,}744+\frac{160{,}256+370L}{R}
\right],
\end{equation}
where the three terms correspond to the input projection,
grouped query-side transformations, and history scan.
The frozen selector incurs only forward computation
when a refresh is required. This represents amortized
user-state maintenance rather than per-candidate
selector execution during online ranking.

The retained-history attention layer uses four heads
with total width 128 and a 256-dimensional target query.
Including query projection, shared key/value projections,
and target-aware scoring and aggregation over $K$ events,
its forward cost is
\begin{equation}
\begin{aligned}
A(K,R)
&=2\cdot256\cdot128
+\frac{4\cdot185\cdot128K}{R}
+4\cdot128K\\
&=65{,}536+\frac{94{,}720K}{R}+512K.
\end{aligned}
\end{equation}
Absorbing the input transformation into the key/value
projections requires a parameter composition cost of
$P=4\cdot185\cdot256\cdot128=24{,}248{,}320$ FLOPs.
These parameters are precomputed for inference but
recomposed during training. With $U$ user groups per
microbatch, the amortized cost is $P/(UR)$ per target,
we set $U=1$.

\paragraph{Training and inference costs.}
We approximate forward and backward computation for
trainable matrix operations as three times the forward
FLOPs, while frozen selector refresh remains forward-only.
The total estimated costs are
\begin{equation}
\begin{aligned}
C_S^{\mathrm{tr}}(L)
&=3C_S(L,40),\\
C_S^{\mathrm{inf}}(L)
&=C_S(L,300),\\
C_E^{\mathrm{tr}}(L)
&=3\left[A(K,40)+\frac{P}{40U}\right]
+S(L,40,0.5),\\
C_E^{\mathrm{inf}}(L)
&=A(K,300)+S(L,300,0.6).
\end{aligned}
\end{equation}
At $L=100{,}000$ and $K=10{,}000$, STCA and ECHO
require 14.262927 and 0.088960 GFLOPs per target
during training, and 1.346437 and 0.008456 GFLOPs
during inference, respectively.
The corresponding ratios are $160.330\times$ and
$159.223\times$, computed from unrounded values. 
For $L>10{,}000$, the attention cost $A(K,R)$
remains fixed, while selector refresh adds only
$(1-h)370/R$ FLOPs per additional historical event.
Thus, ECHO's estimated cost grows slowly with $L$.

\vspace{-0.5em}
\subsection{Additional Experimental Results}
\label{app:additional_results}

\begin{figure}[t]
 \centering
 \scalebox{\FigEightScale}{\includegraphics[width=\linewidth]{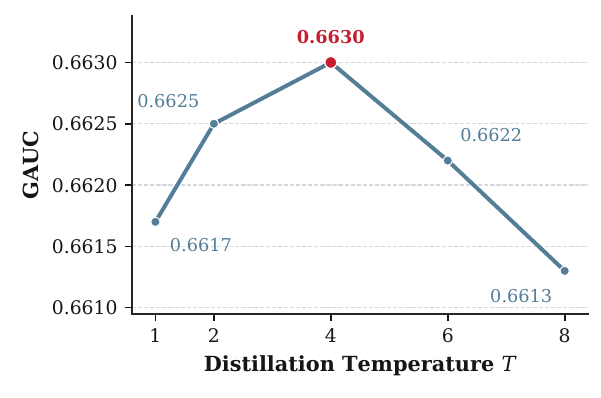}}
 \caption{Effect of distillation temperature on ECHO's Click GAUC.}
 \Description{Click GAUC at temperatures 1, 2, 4, 6, and 8 is 0.6617, 0.6625, 0.6630, 0.6622, and 0.6613, respectively. The highest observed GAUC occurs at temperature 4.}
 \vspace{-1.5em}
 \label{fig:temperature_ablation}
\end{figure}

\begin{figure}[t]
 \centering
 \scalebox{\FigNineScale}{\includegraphics[width=\linewidth]{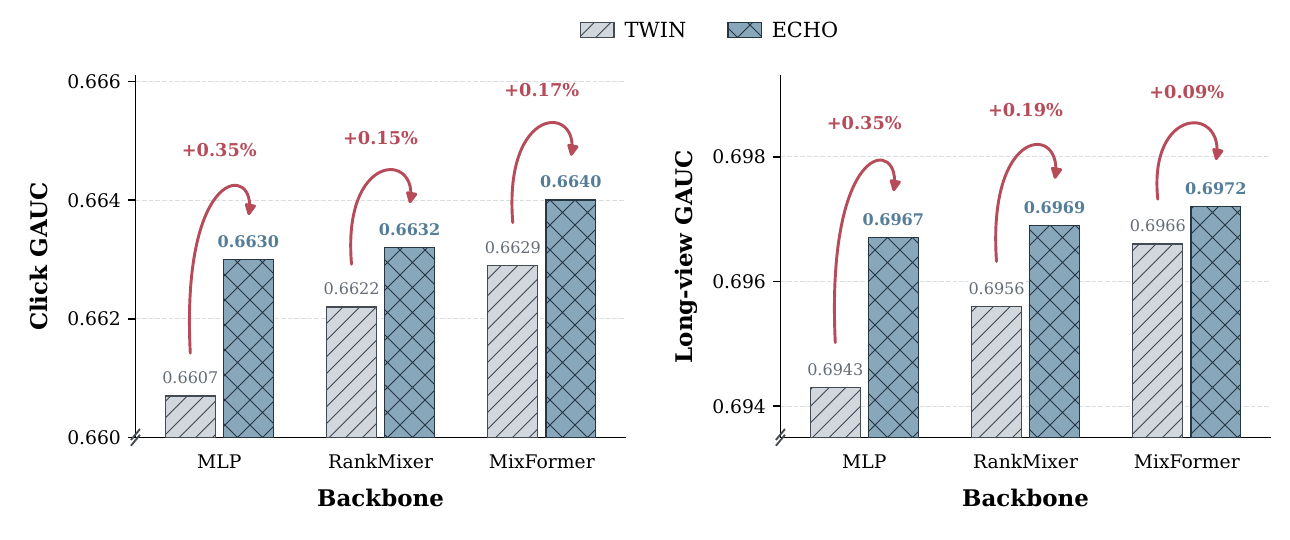}}
 \caption{Model-agnostic evaluation on KuaiRand-1K. Click and Long-view GAUC of TWIN and ECHO with different downstream backbones. Curved arrows indicate relative GAUC improvements over TWIN.}
 \Description{Grouped bars compare TWIN and ECHO with MLP, RankMixer, and MixFormer on Click and Long-view. ECHO has higher GAUC in all six comparisons. Curved arrows show relative improvements, and breaks mark the nonzero GAUC axis baselines.}
 \label{fig:model_agnostic}
 \vspace{-1.5em}
\end{figure}

We investigate the sensitivity of ECHO to the distillation temperature \(T\), which controls the smoothness of the teacher's relevance distribution and consequently affects the supervision provided to the selector. We vary \(T\) over \(\{1,2,4,6,8\}\) and evaluate Click GAUC on KuaiRand-1K. As illustrated in Figure~\ref{fig:temperature_ablation}, GAUC increases from 0.6617 at \(T=1\) to 0.6630 at \(T=4\), before decreasing to 0.6613 at \(T=8\). This indicates that the sharpness of the distillation distribution plays an important role in learning effective behavior selection. A lower temperature concentrates supervision on a small number of highly relevant behaviors, potentially overlooking useful relevance information from other interactions. Conversely, an excessively high temperature produces a smoother distribution that may weaken the distinction between behaviors of different relevance levels. An intermediate temperature provides a better balance between emphasizing highly relevant behaviors and preserving richer relevance information, leading to improved prediction performance.

Finally, we evaluate the applicability of ECHO across different downstream ranking architectures. Since ECHO performs user-level history selection while preserving original behavior records, its selection mechanism can be integrated with different feature interaction networks. We compare ECHO with TWIN using three downstream backbones: MLP, RankMixer, and MixFormer. As shown in Figure~\ref{fig:model_agnostic}, ECHO consistently outperforms TWIN across all three architectures, achieving relative GAUC improvements of 0.15\%–0.35\% on Click and 0.09\%–0.35\% on Long-view. In particular, ECHO improves Click GAUC from 0.6607 to 0.6630 with MLP and from 0.6629 to 0.6640 with MixFormer. Despite variations in improvement across backbones, the consistent gains across all six comparisons indicate that ECHO can provide informative historical evidence for different downstream models, demonstrating its flexibility and model-agnostic applicability.

Overall, these experiments validate the effectiveness of distillation-based supervision, highlight the importance of an appropriate distillation temperature, and demonstrate that ECHO can effectively complement diverse downstream ranking architectures.

\section{Limitations} \label{app:limitations}

\paragraph{Long-tail limitations.}
Compressing 100K events to 10K may discard history relevant to extremely long-tail targets, making it unavailable to downstream retrieval. Unlike SequenceO1~\cite{guan2026sequenceo1}, which compresses sequences into tokens, ECHO retains original records, allowing direct access to item identities, timestamps, and side features when relevant events survive selection and are retrieved. However, long-tail items can still be discarded during compression.

\paragraph{Adaptation latency.}
The selector adapts through training, not to individual online targets. Recovering excluded events for long-tail targets requires relevant training signals, model updates, and selection refreshes.

\end{document}